\documentstyle[12pt,epsfig,rotating]{article}
\newcommand{\ra}{\rightarrow}
\newcommand{\ftc}{F_2^{c\bar{c}}}

\begin{document}
\topskip 2cm 
\begin{titlepage}


\begin{center}
{\large\bf HEAVY FLAVOUR PRODUCTION AT HERA} \\
\vspace{2.5cm}
{\large Felix Sefkow} \\
\vspace{.5cm}
{\sl Physik-Institut der Universit\"at Z\"urich}\\
{\sl CH-8057 Z\"urich, Switzerland}\\
\vspace{2.5cm}
\vfil
\begin{abstract}

Recent results from the experiments ZEUS and H1
on charm production in $ep$ collisions 
are reviewed. 
The topics are elastic and inelastic $J/\psi$ photoproduction, 
$D^*$ photoproduction differential cross sections and 
a first look at the proton structure function 
$\ftc$. 
 
\end{abstract}

\vspace{4cm}
{\it Invited talk presented at
Les Rencontres de Physique de la Vall\'ee d'Aoste: \\ 
Results and Perspectives in Particle Physics,
La Thuile, Italy, March 3-9, 1996,\\
on behalf of the ZEUS and H1 collaborations} 

\end{center}.
\end{titlepage}

\section{Introduction}

The production of heavy quarks in electron proton interactions
proceeds, in QCD, almost exclusively via photon gluon fusion, where 
a photon emitted from the incoming electron interacts 
with a gluon in the proton by forming a quark-antiquark pair.
Therefore, heavy quarks, in particular charm, 
offer the classical way~\cite{emc_openc,nmc_psi} 
towards a determination of the gluon density in the proton.

Beauty production, with respect to charm, is expected~\cite{eichler}
to be suppressed by about two orders of magnitude:
at HERA, 
for $\sqrt{s}=300$ GeV, 
cross sections are  
$\sigma(ep\ra c\bar{c}) \approx 1 \mu$b and  
$\sigma(ep\ra b\bar{b}) \approx 5 $nb.
No indication of $b$ or $\Upsilon$ production has been reported
by neither the H1 nor ZEUS collaboration up to now. 
With the total luminosity accumulated so far (10 pb$^{-1}$),
the era of HERA $b$ physics has not yet started.    
(Truth production will never happen at HERA, obviously.)

Consequently, this talk deals with charm only,
with the production of hidden charm in the first, with open charm
in the second part.     
In section 2, elastic and inelastic $J/\psi$ production are 
discussed, in section 3 new results on $D^*$ photoproduction
are presented, and a first measurement of the proton 
structure function $\ftc$ is shown.

\section{Hidden Charm}

{\bf HERA Kinematics:}
At HERA, electrons (or positrons) of energy 27 GeV 
collide head-on with 820 GeV protons,
providing a center-of-mass energy $\sqrt{s}=$ 300 GeV.
The kinematics of deep inelastic scattering (DIS) is described 
using the well-known Lorentz invariants 
$Q^2 = -q^2$ and 
$x = Q^2 / (2pq)$, 
where $p$ is the 4-momentum of the incoming proton 
and $q$ that of the exchanged photon. 
The scaling variable $y$ is related to these by
$Q^2=xys$.
In the photoproduction limit 
$Q^2 \approx 0$, 
as in the DIS case,   
typical photon-proton center-of-mass energies $W$ are  
100 - 200 GeV.
HERA thus extends the range of fixed-target photoproduction 
experiments by an order of magnitude. 
Consequently, given the same scale of the partonic subprocess,  
the parton densities in the proton are probed
at much smaller momentum fractions
of the parton in the proton,
$x_p \sim 10^{-3}$.

\noindent {\bf Classification:}
To the photoproduction of $J/\psi$ mesons, 
several processes contribute which lead to 
well distinguished experimental signatures.
In the elastic process 
$\gamma p\ra J/\psi\, p$,
the proton stays intact and 
leaves the detector through the beam-pipe,
such that nothing but the $J/\psi$ decay products is detected.
The scattering is diffractive, i.e.\ 
only 4-momentum is transferred to the proton, 
no color is exchanged.
Diffractive processes where the proton breaks up 
("diffractive $p$-dissociation") 
constitute a background to the exclusive elastic reaction. 
It can however effectively be suppressed 
by rejecting events where apart from the $J/\psi$, 
proton debris is detected under very small angles. 

In contrast, in the inelastic case 
a hard gluon from the proton interacts with the 
photon to form the $c\bar{c}$ pair. 
According to the color singlet model~\cite{coloursing}
another gluon has to be subsequently radiated, in order 
to restore the $J/\psi$ color quantum numbers.      
The inelastic process involves color flow 
between the proton and the charm system, and after hadronization 
energy depositions are spread over wider regions of the detector. 

The processes may also be separated by use 
of the elasticity variable $z$ that can be calculated 
from the longitudinal momenta of the $J/\psi$ products
and of all final state particles: 
$z = (E-p_z)_{J/\psi}/(E-p_z)_{\mbox{\small all}}$.
Elastic production gives $z \approx 1$, 
for diffractive dissociation $z$ is smaller, 
but still close to 1.
Inelastic production leads to values $0<z<1$. 
There is also a so-called resolved contribution, 
where the photon fluctuates into a hadronic state and 
a parton from that state interacts with the proton. 
These processes cluster at small $z$ and can thus be
suppressed by a lower cut on this variable.

\subsection{Elastic $J/\psi$ production}

\begin{figure}[tb]\centering
\vspace{-5cm}
\begin{picture}(470,630)(0,-200)
\epsfig{file=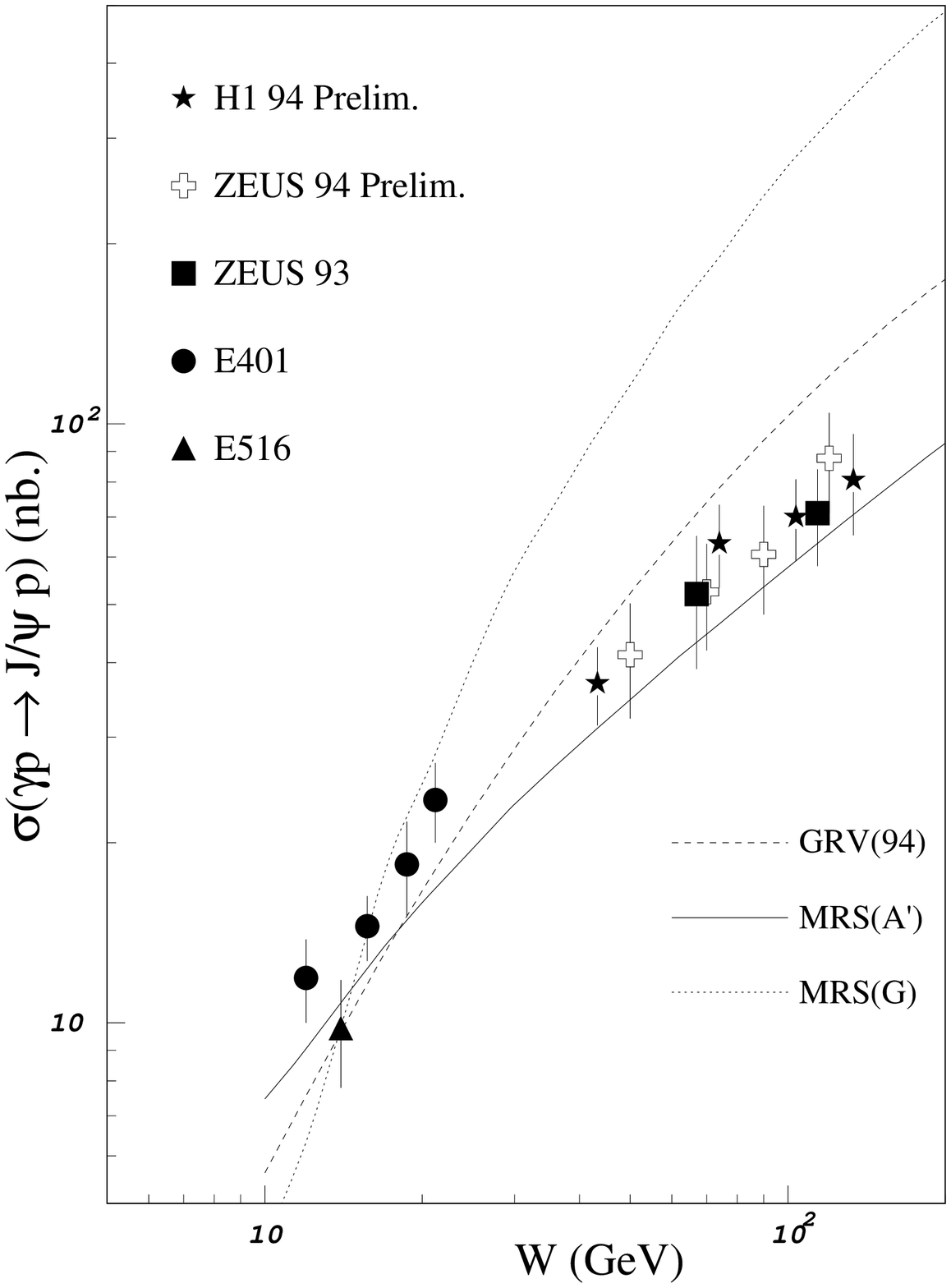,height=9cm}
\end{picture}
\begin{picture}(370,360)(-160,-560)
\epsfig{file=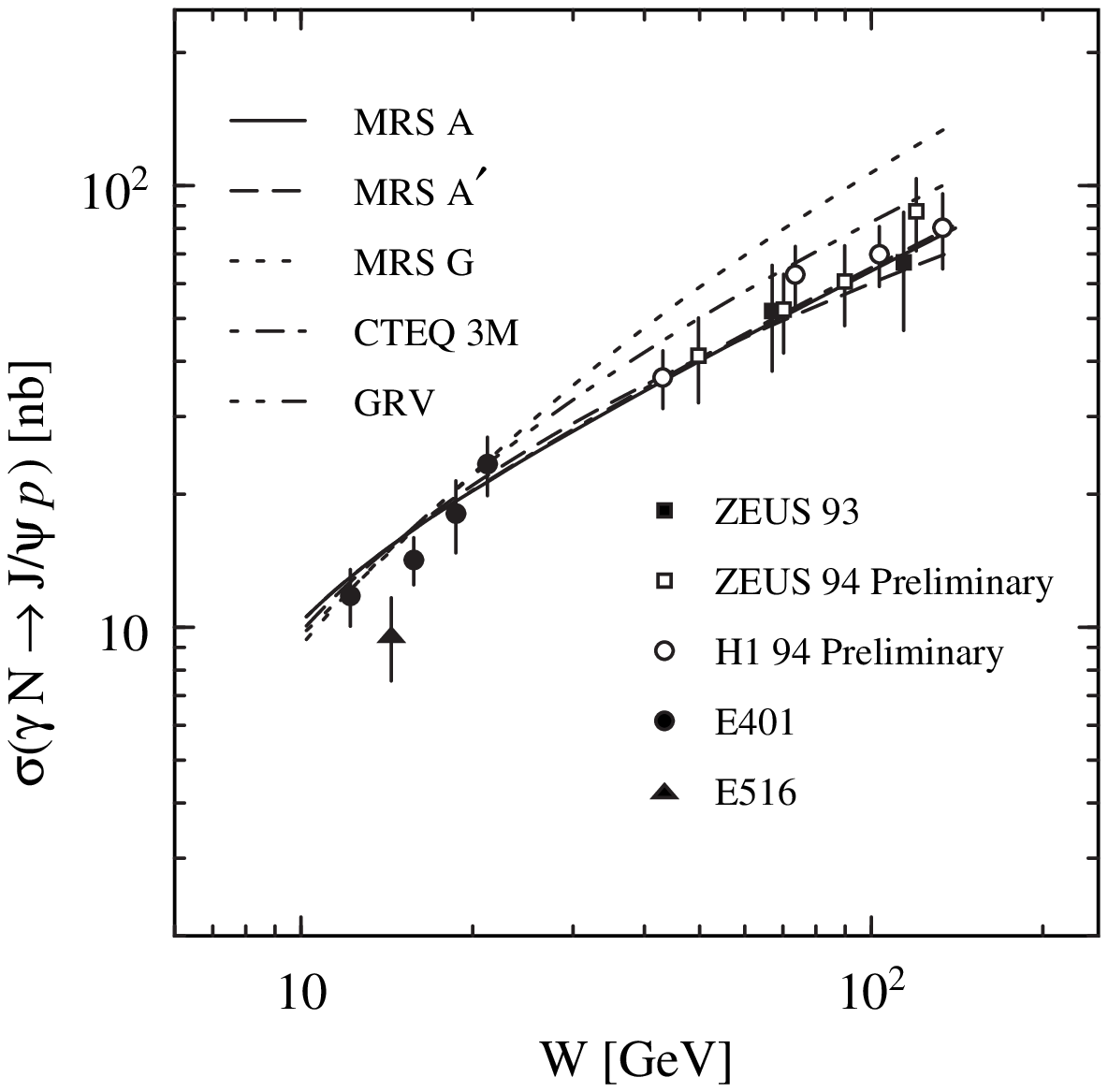,height=9cm}
\end{picture}
\begin{picture}(1,1)
\put(-290,792){(a)}
\end{picture}
\begin{picture}(1,1)
\put(-60,780){(b)}
\end{picture}
\vspace{-20cm}
\caption[dummy]%
  {Elastic $J/\psi$ photoproduction cross section, 
   measured at fixed target and HERA experiments,
   versus $\gamma p$ center-of-mass energy.  
   The curves are the QCD predictions of the 
   gluon ladder model~\cite{ryskin} (a)
   and of the color octet model~\cite{fleming} (b).} 
\label{fig:elpsi}
\end{figure}

The ZEUS and H1 experiments have meanwhile --
with integrated luminosities of about 3 pb$^{-1}$ collected in 94 -- 
accumulated samples of order 1000 elastic events each,
where the $J/\psi$ mesons are reconstructed in the decay channels 
$J/\psi \ra \mu^+\mu^- , e^+ e^-$. 
The measured cross sections are corrected for remaining contaminations 
from diffractive dissociation 
($(17^{+8}_{-5}\pm 10)\%$ and $(12\pm12)\%$ for ZEUS and H1,
respectively) 
and for feed-down from $\psi^{\prime}$ decays. 
The results~\cite{elpsizeus,psih1}
 are displayed as a function of the $\gamma p$ 
center-of-mass energy $W$ in Fig.~\ref{fig:elpsi}a
together with results from fixed-target experiments~\cite{FTelpsi}. 
The cross section exhibits a steep rise that cannot be explained
in the framework of a soft vector meson dominance model~\cite{landshoff} 
where an energy dependence 
$\sigma_{el} \sim W^{4\epsilon}$ with $\epsilon \approx  0.08$
is expected. This conclusion can be drawn from the HERA data alone. 

Ryskin and co-workers have argued~\cite{ryskin1}
that due to the high scale of the process given by 
$Q_0^2=M^2_{J/\psi}/4$, 
perturbative QCD methods should be applicable. 
In their model, a gluon ladder diagram gives 
the dominant contribution to the cross section, 
which is therefore found to be proportional 
to the {\it square} of the gluon density in the proton: 
$\sigma_{el} \sim [ g(x)]^2$ with $x = M^2_{J/\psi}/W^2\sim 10^{-3}$.
Recent leading order (LO)
calculations
in the Ryskin model,
including some next-to-leading (NLO) effects~\cite{ryskin},  
are also displayed in Fig.~\ref{fig:elpsi}a;
using different parameterizations  
of the gluon density in the proton~\cite{mrs,grv}, 
which are all consistent with inclusive structure function measurements
at HERA~\cite{f2h1zeus}.
The model can reproduce the steep rise of the cross section with $W$. 
The potentially high sensitivity of the elastic $J/\psi$ cross section
as a probe of the gluon density is clearly illustrated. 
However,
before this can be used as a measurement of $g(x)$,
a better understanding of the model uncertainties has to be gained. 

A complementary approach is provided by the non-relativistic QCD 
(NRQCD) scheme developed by 
Bodwin, Braaten and Lepage~\cite{NRQCD}.  
Here, the $J/\psi$ production amplitude is factorized into a 
hard boson-gluon fusion part  -- 
yielding a $c\bar{c}$ system in a color octet state in the first place -- 
and a soft part that describes the subsequent transition 
of the octet state into a color singlet $J/\psi$ meson. 
The latter transition is described by color octet matrix elements,
the cross section is symbolically written as
$\sigma \sim g(x)\cdot 
\sum_{[n]} \langle 0|{\cal O}_{\underline{8}} [n] | 0\rangle$,
where $[n]$ runs over the dominating angular momentum states. 
The matrix elements are treated as phenomenological parameters;
fitting them to lower energy data
gives a prediction~\cite{fleming} for the HERA regime.
This is shown in Fig.~\ref{fig:elpsi}b together 
with the same data as in Fig.~\ref{fig:elpsi}a.
Again, a good description of the measurements is obtained, 
using, for example, the parton density parameterization set 
MRS(A')~\cite{mrs}.
Note, however, that in this model the cross section depends 
{\em linearly} on the gluon density $g(x)$. --
The theoretical debate on how to best describe
elastic $J/\psi$ photoproduction is open at present.

\subsection{Inelastic $J/\psi$ production}

\begin{figure}[tbp]\centering
\vspace{2cm}
\epsfig{file=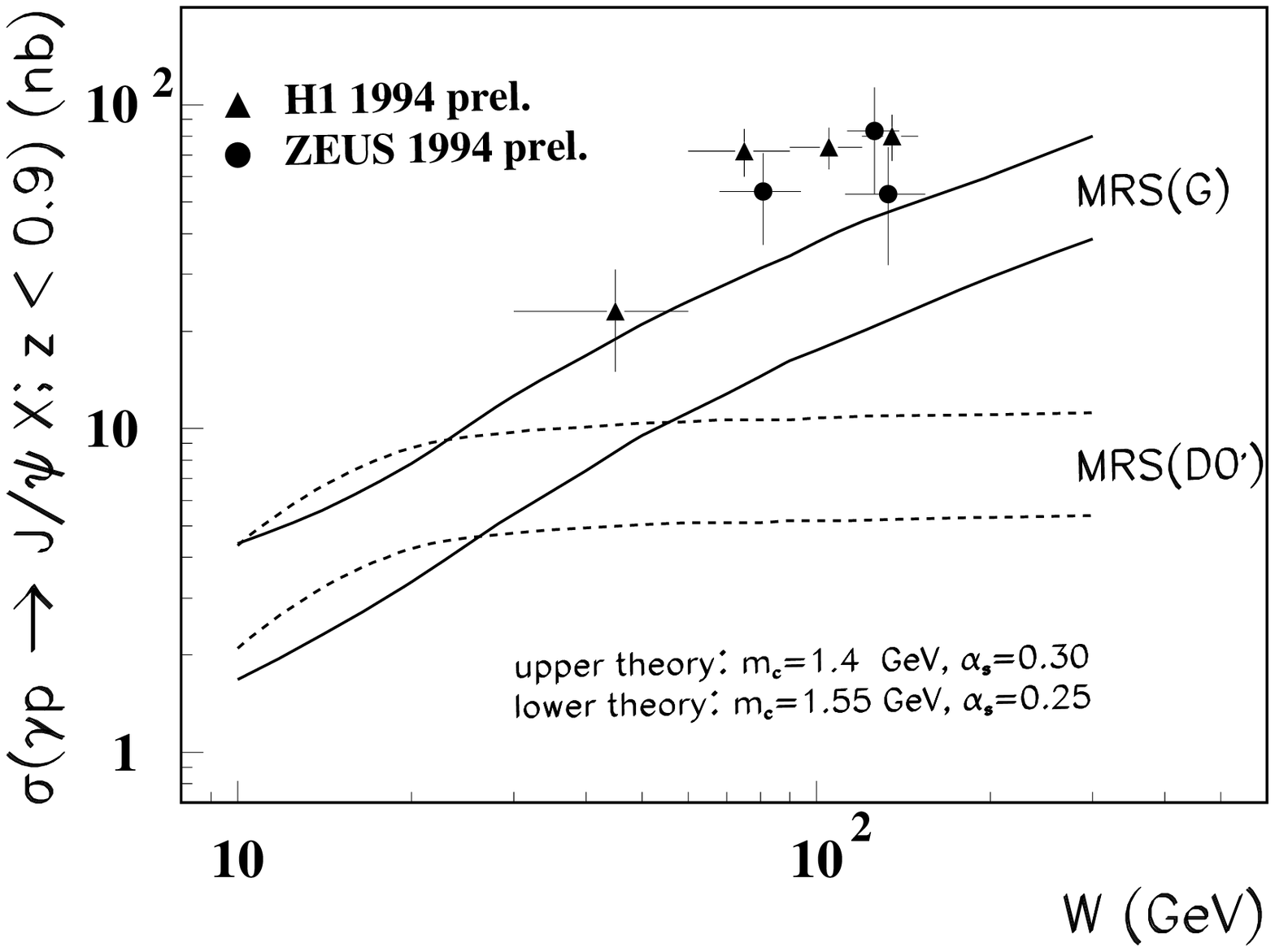,height=6cm}
\epsfig{file=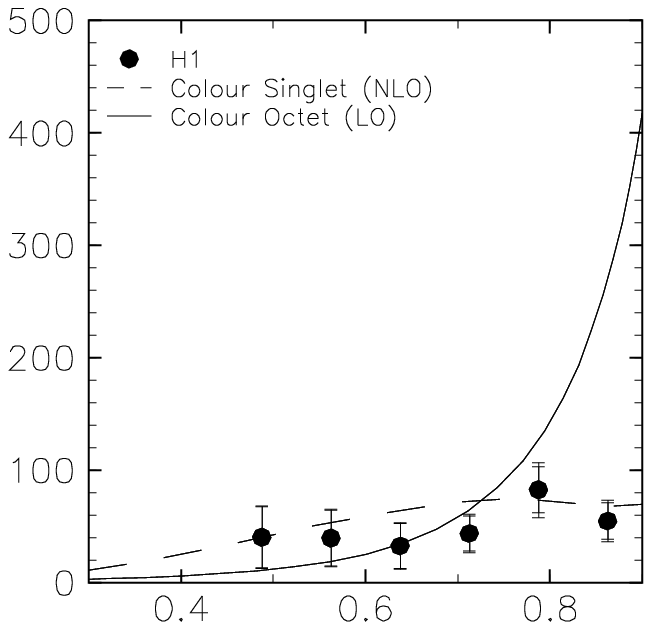,height=7cm,angle=90}
\begin{picture}(20,20)
\put(185,30){$z$}
\end{picture}
\begin{picture}(20,20)
\put(45,60){\begin{rotate}{90}
$d\sigma/dz(\gamma p \ra J/\psi\, X)(\mbox{nb})$\end{rotate}}
\end{picture}
\begin{picture}(1,1)
\put(-37,172){(a)}
\end{picture}
\begin{picture}(1,1)
\put(162,171){(b)}
\end{picture}


\caption[dummy]%
        {a) Inelastic $J/\psi$ photoproduction 
        cross section versus $W$. 
        The curves are NLO QCD predictions for 
        two different sets of parton density parameterizations. 
        b) Inelastic $J/\psi$ cross section 
        versus elasticity $z$, 
        together with QCD calculations
        of color singlet and possible octet contributions 
        (see text).}
\label{fig:inelpsi}
\end{figure}

Cross section measurements done by ZEUS~\cite{inelzeus}
and H1~\cite{psih1}
for inelastic $J/\psi$ production in the range $z<0.9$ 
are shown as a function of $W$ in Fig.~\ref{fig:inelpsi}a
and compared to next-to-leading order 
QCD calculations~\cite{kraemer}. 

The measurements are done at $z$ above 0.4 typically 
and extrapolated to $z=0$; the correction is about 10\%.
The agreement with NLO QCD is best when a steeply rising 
gluon density distribution (like MRS(G)~\cite{mrs})
is used,
together with a low value of the charm quark mass
and a rather large strong coupling constant.
Nevertheless, the NLO calculation still falls short 
in describing the normalization of the data
by some amount,
indicating that contributions from even higher orders may be significant.
The calculation is considered most reliable in the region where 
the NLO corrections are small, this is for $z<0.8$ and transverse 
momenta of the $J/\psi$ $p_{\perp}>1$ GeV/c. 
The H1 measurements in this restricted range 
are in very good agreement with NLO QCD,
but the sensitivity to the gluon density in the proton 
is much reduced.

The $z$ dependence of the cross section is sensitive to possible
color octet contributions to $J/\psi$ production. 
Such processes have recently been proposed~\cite{octet_tev}
to explain the high quarkonium production rates observed at the Tevatron.
The differential cross section $d\sigma /dz$ has been measured  by 
H1 and is shown in Fig.~\ref{fig:inelpsi}b,
together with calculations~\cite{cacciari} 
drawn separately 
for the familiar color singlet 
contributions and for color octet contributions.
The color octet matrix elements used in the calculation 
were extracted from fits 
to prompt $J/\psi$ production data from the Tevatron~\cite{psi_tev}.
The measured distribution clearly disfavors large color octet 
contributions to inelastic $J/\psi$ production at HERA.

\section{Open Charm}

\subsection{$D^*$ photoproduction}

Photoproduction of open charm occurs via direct photon gluon fusion, 
$\gamma g \ra c\bar{c}$ or via resolved processes where a parton 
inside the photon scatters off a parton inside the proton, e.g.\ 
$gg\ra c\bar{c}$. 
The latter are known to dominate the production of light quarks,   
but are expected to contribute much less to heavy quark production. 
Due to the smaller available energy of the parton from the photon side, 
and the consequently lower center-of-mass energy of the hard subprocess, 
the resolved events are characterized by a stronger boost into the 
proton direction and smaller transverse momenta $p_{\perp}$.
The experimental cuts 
limit the measurement to central rapidities and large $p_{\perp}$
and thus additionally suppress the resolved contribution
to a level of below 10 \% typically.

Both ZEUS and H1 have tagged $c\bar{c}$ events through the detection 
of muons from semileptonic decays and through reconstruction 
of  $D^*$ decays, and derived cross sections from the two methods. 
The published results~\cite{dstzeus,dsth1} that 
take advantage of the well known clean signature of the decay chain 
$D^{*+}\ra D^0\pi^+$, $D^0\ra K^-\pi^+$
are more precise than 
the so far presented (preliminary)
inclusive muon data which suffer from higher background~\cite{inclmu}. 

\begin{figure}[tbp]\centering
\epsfig{file=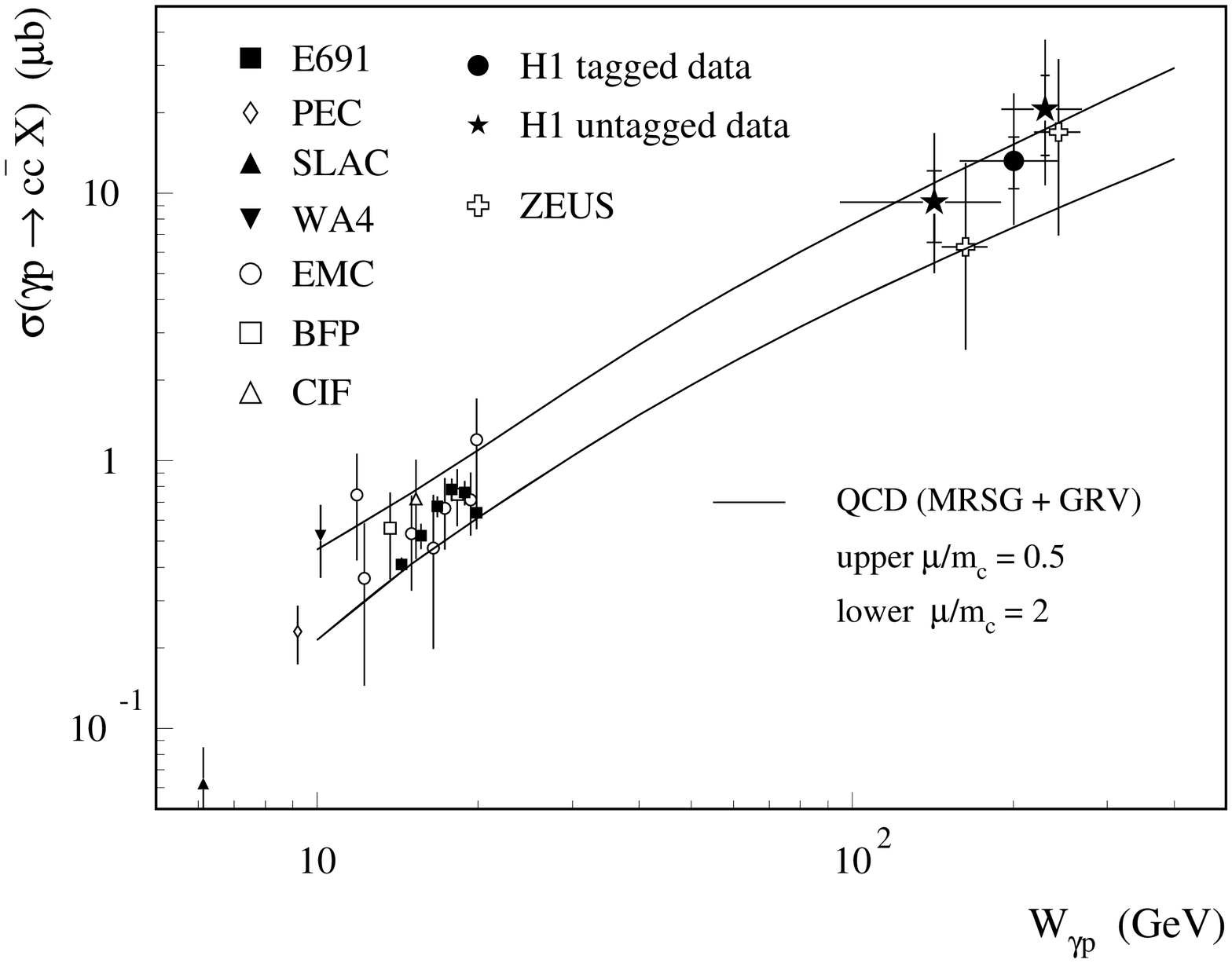,width=8.5cm}
\epsfig{file=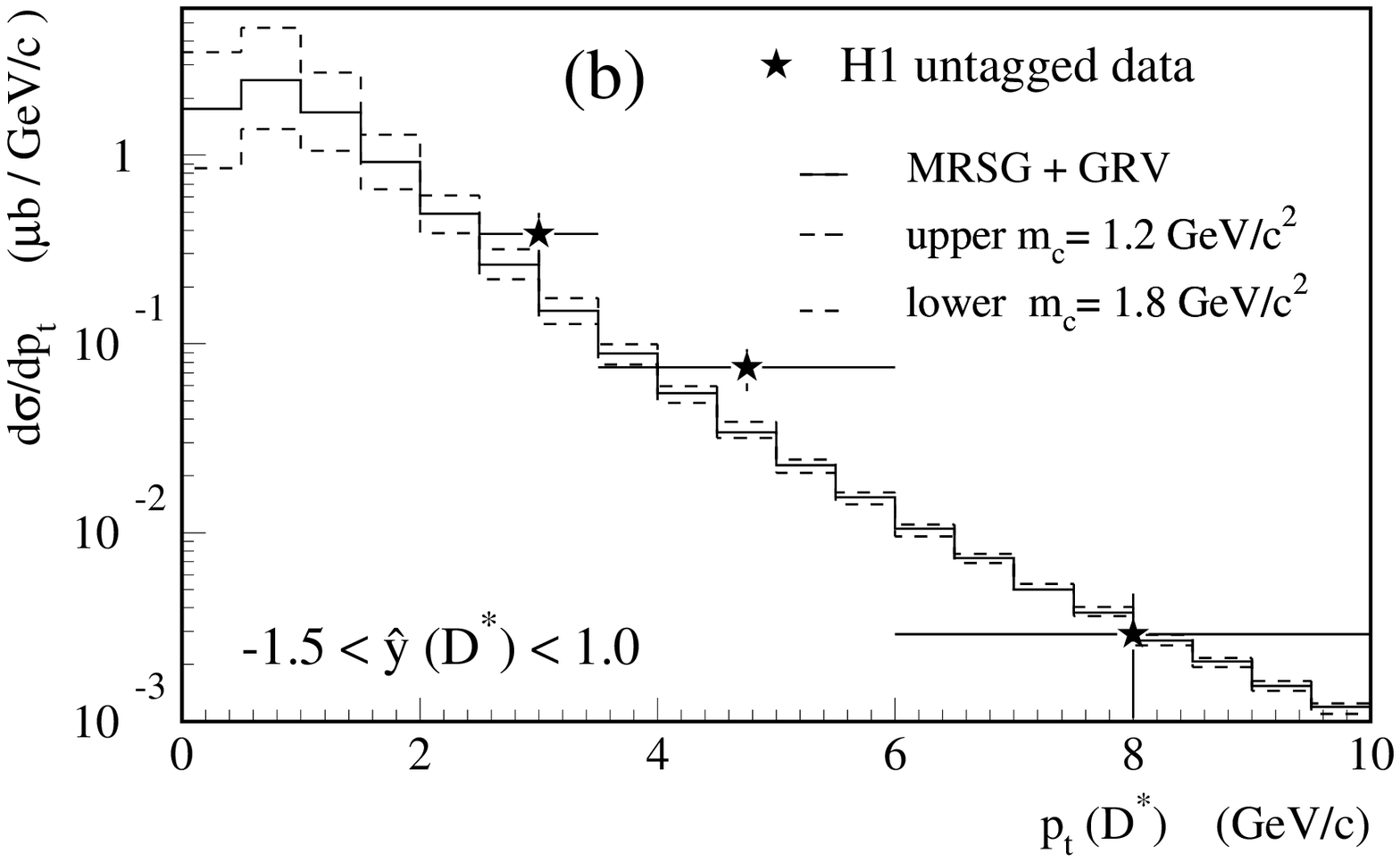,width=9cm}
\epsfig{file=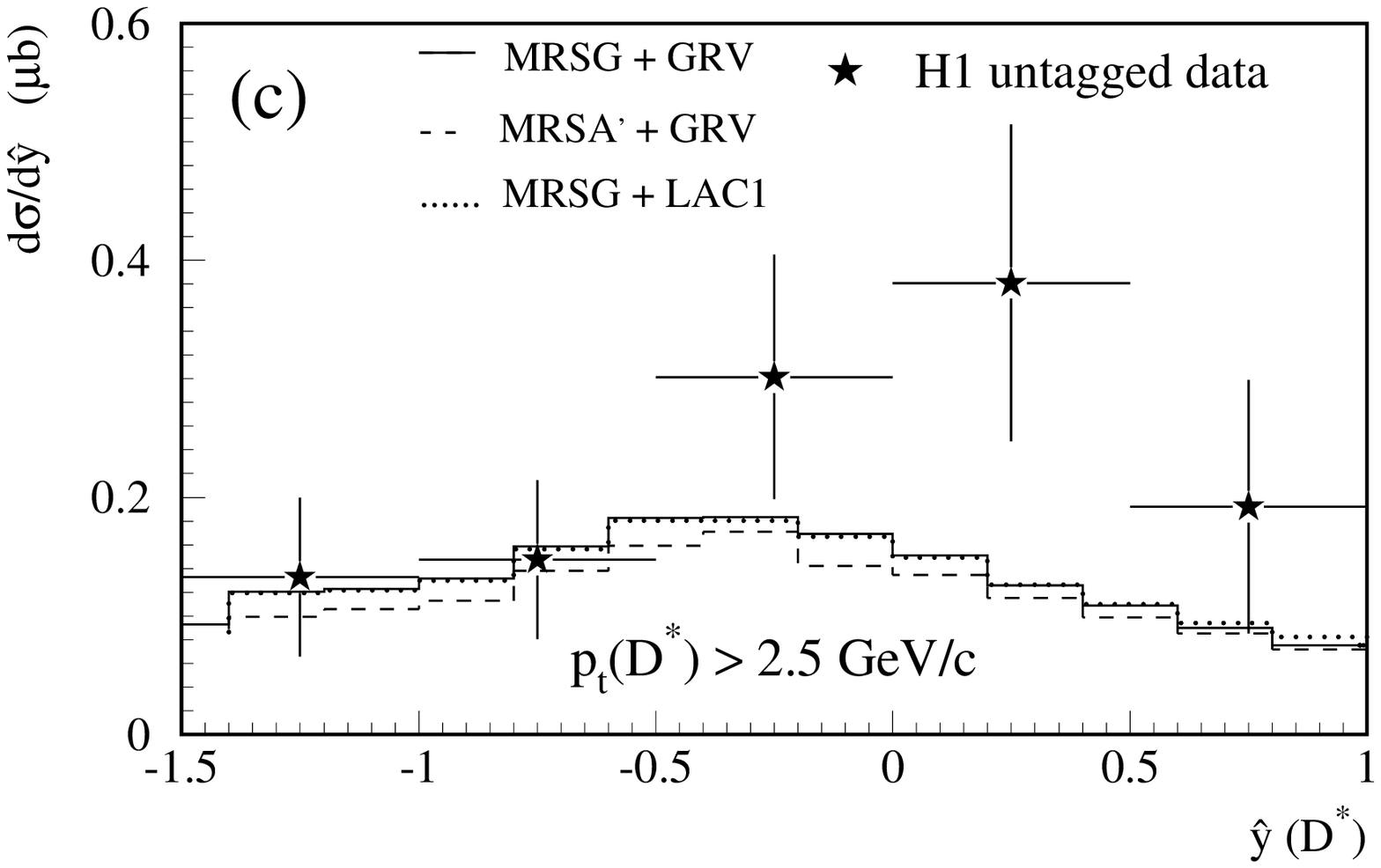,width=9cm}
\begin{picture}(1,1)
\put(-150,345){(a)}
\end{picture}
\caption[dummy]%
        {a) Total $\gamma p \ra c\bar{c}X$ cross section 
            versus $\gamma p$ center-of-mass energy $W$.
            The curves are NLO QCD predictions
            for different factorization and renormalization 
            scales $\mu$. 
         b) Differential charm cross section versus $p_{\perp}$.
            The histograms are NLO QCD predictions for different 
            values of the charm quark mass.
         c) Differential charm cross section versus rapidity.
            The histograms are NLO QCD predictions for
            different parameterizations of the parton 
            densities, showing the effect of varying the 
            proton (dashed vs.\ full)
            or photon (dotted vs.\ full) densities.}
\label{fig:dstgp}
\end{figure}

The new H1 analysis of 94 data exploits
a $D^*$ signal of more than 200 events
in the kinematical region $p_{\perp}(D^*) >2.5$ GeV/c and rapidity 
$-1.5<\hat{y}<1$.
The measurement is done for the case 
where the scattered positron is either registered in the electron detector  
of the luminosity system ("tagged"), or not required to be seen  
at all ("untagged").
The total charm cross section is shown as a function of 
the $\gamma p$ center-of-mass energy $W$ 
in Fig.~\ref{fig:dstgp},
together with the ZEUS result (using 93 data)~\cite{dstzeus}
and previous fixed target measurements~\cite{FTgpcc}.
The cross section rises strongly with energy; 
at HERA it is about an order of magnitude higher than 
at fixed target energies. 

There are however large extrapolation factors involved in
the transformation from the visible $p_{\perp}$ and $\hat{y}$ range  
to total cross sections. 
These give rise to large systematic errors associated with 
uncertainties in the parton distributions in the proton and 
the photon.
They are included in the total errors in Fig.~\ref{fig:dstgp}a,
the inner error bars of the H1 data 
indicate the experimental errors alone. 
The following example may illustrate 
the dependences.
The H1 measurement with tagged data, at $W=200$ GeV, 
would change from $(12.2 \pm 2.0 \pm 1.8)\,\mu$b 
to $(7.4 \pm 1.2 \pm 1.1)\,\mu$b, 
if the parton density parameterization 
MRS(D0') instead of the steeper MRS(A') set~\cite{mrs}
were used for the extrapolation. 
However, the QCD prediction changes in the same direction:
from $9.8\,\mu$b to $3.9\,\mu$b, respectively. 
A similar picture is obtained for variations of the 
parton densities in the photon. 
Hence the total cross section is not well suited for
the determination of gluon densities.  

{\bf Differential cross sections:}
H1 has therefore returned to the visible kinematic 
range that is free of extrapolation uncertainties 
and measured differential cross sections~\cite{dsth1}. 
The $p_{\perp}$ distribution is shown in Fig.~\ref{fig:dstgp}b
and compared to NLO QCD calculations using the program of
S.\ Frixione {\it et al}.~\cite{frixione_diff}. 
The data are in good agreement with QCD. 
The figure illustrates that the value for the charm mass    
used in the calculation mainly affects the region of low 
$p_{\perp}<3\,$GeV. -- 
The rapidity distribution is displayed in Fig.~\ref{fig:dstgp}c
together with the  results of NLO QCD calculations
where the gluon density has been varied in the proton (dashed
histogram), or in the photon (dotted).
The backward region of negative rapidities is most 
sensitive to the parton content of the proton,
whereas the effect of the photon structure is 
most pronounced in the forward direction, outside the 
visible range shown.   
The agreement between data and theory is only marginal here.
This may give rise to speculations about possible 
contributions from the excitation of intrinsic charm 
in the photon~\cite{godbole};  
the experimental errors are however still too large to 
draw firm conclusions.  

\begin{figure}[tbp]\centering
\vspace{1cm}
\epsfig{file=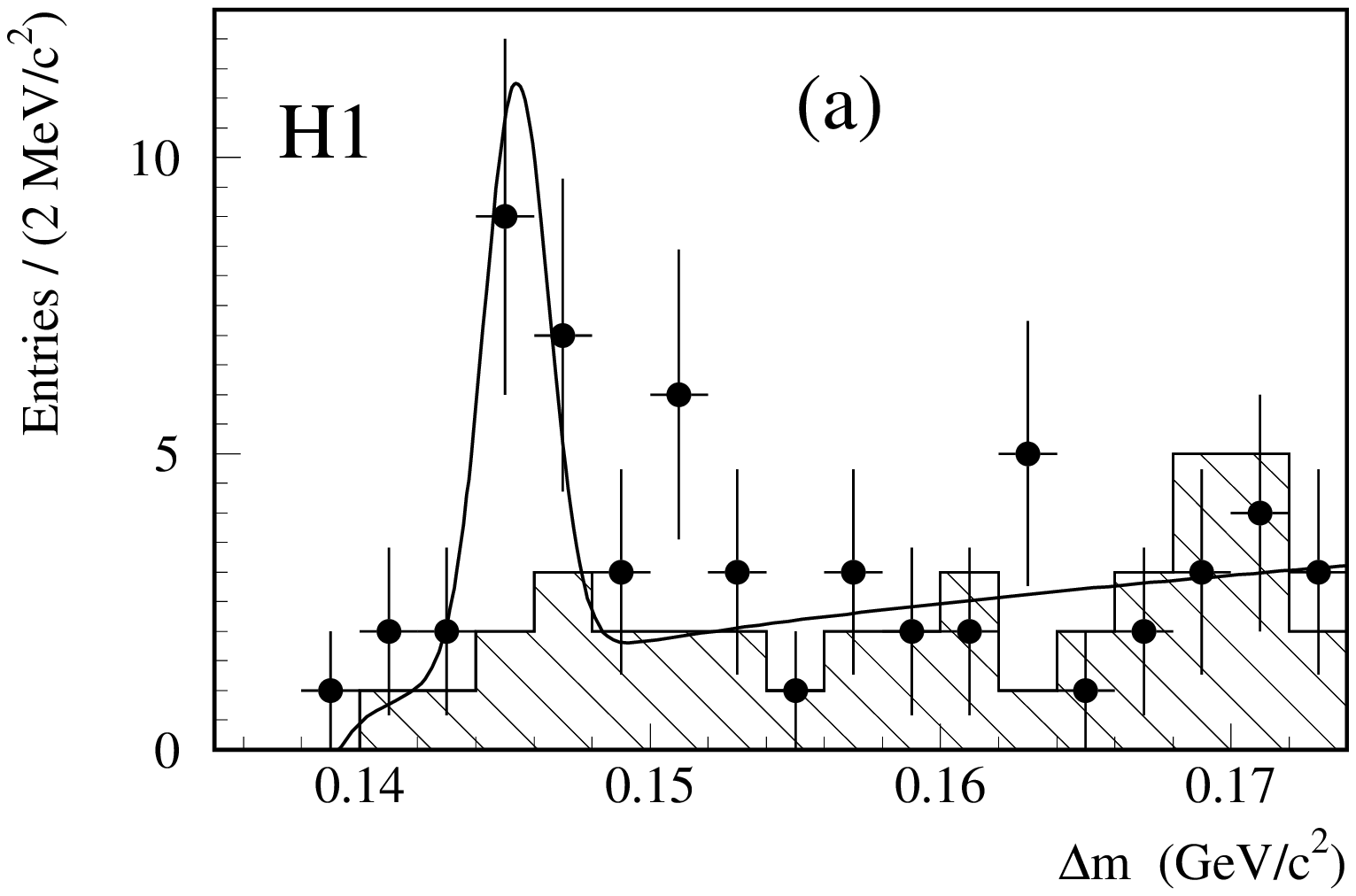,height=5.5cm}
\epsfig{file=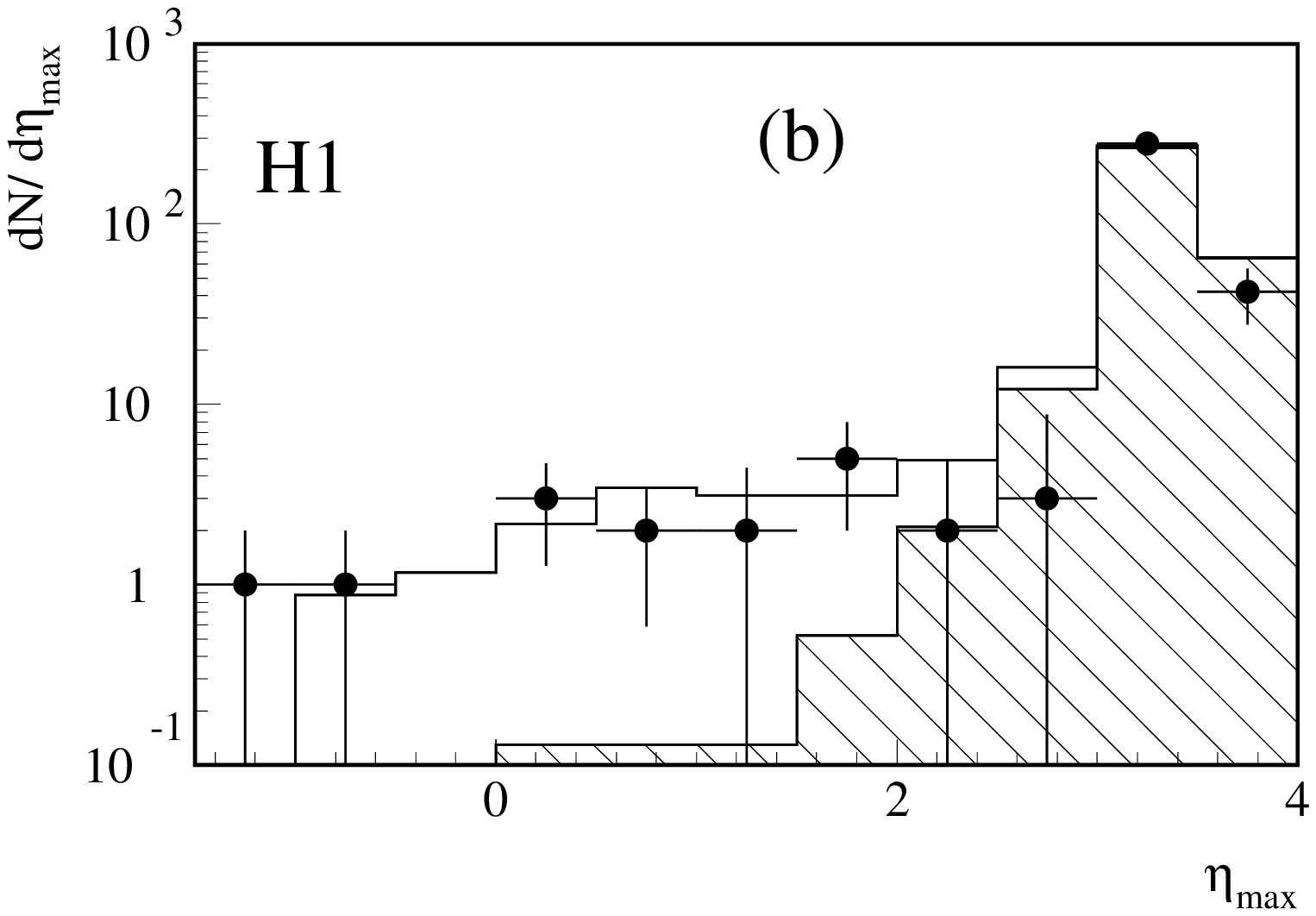,height=5.5cm}
\caption[dummy]%
        {a) $D^*$ signal in the invariant mass difference 
         $\Delta M = M(K\pi\pi ) - M(K\pi )$ for photoproduction 
         events with a rapidity gap. 
         The hatched histogram indicates the background estimated 
         from wrong charge combinations. 
         b) Background subtracted $\eta _{max}$ distribution
         of $D^*$ events.
         The histograms are Monte Carlo predictions (hatched:
         PYTHIA~\cite{pythia}, open: RAPGAP~\cite{rapgap})
         normalized to the data (see text).}
\label{fig:dstdiff}
\end{figure}

\noindent {\bf Diffractive production:}
H1 has found first evidence for diffractive charm photoproduction 
at HERA,
by searching for $D^*$ mesons in (untagged) $\gamma p$ events with a 
rapidity gap.
The selection requires 
the pseudo-rapidity of the most forward calorimetric energy deposition
to be $\eta_{max}<2$, 
as in~\cite{f2dh1}. 
A clear signal is found in the mass difference distribution 
of $\Delta M = M(K\pi\pi ) - M(K\pi )$ for such events, 
see Fig.~\ref{fig:dstdiff}a.
In order to substantiate the evidence for a diffractive production 
mechanism further, the background subtracted $\eta_{max}$ 
distribution of the $D^*$ events is shown in Fig.~\ref{fig:dstdiff}b.
The standard PYTHIA~\cite{pythia} 
Monte Carlo prediction (hatched), 
normalized to the data with $\eta_{max}>2$, 
clearly fails to reproduce the tail in the data towards low
$\eta_{max}$. 
The open histogram is a prediction of the 
RAPGAP Monte Carlo program~\cite{rapgap}
(normalized at $\eta_{max}<2$),
where a partonic structure of the diffractive exchange 
is assumed. 
In this model, the magnitude of the observed signal clearly disfavors a 
quark-dominated exchange, whereas it is consistent with a gluon-dominated
structure.  

\subsection{Open charm in DIS: $\ftc$}

Charmed mesons have also been observed by H1~\cite{dsdh1eps} 
and ZEUS
in deep inelastic scattering.
Here the outgoing electron is measured in the central detectors and   
the kinematics ($x$, $Q^2$) of the scattering process 
is determined from the measured electron energy and direction.  
In addition to the $D^*$ channel measured also in photoproduction,
a $D^0$ signal from the decay $D^0\ra K^-\pi^+$ has been 
seen in the $M(K^-\pi^+)$ invariant mass distribution.
For $Q^2 > 10$ GeV$^2$, H1 quotes from their 94 data 
inclusive cross sections%
\footnote{Charge conjugate states are implicitly included.}
of
$\sigma (ep\ra eD^{*+} X) = (9.6 \pm 1.1 \pm 1.3)\,$pb and  
$\sigma (ep\ra eD^0 X) = (22.5 \pm 3.6 \pm 2.9)\,$pb
the ratio of the two being about as expected. 
    
\begin{figure}[tbp]\centering
\epsfig{file=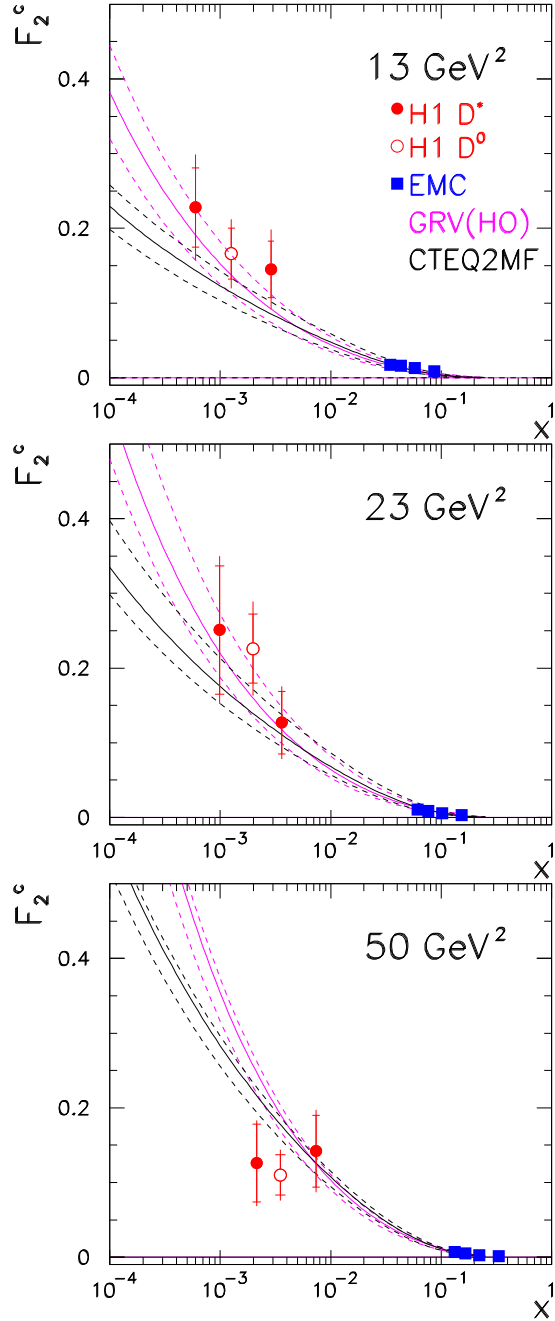,height=18cm}
\caption[dummy]%
        {Charm contribution $\ftc$ to the proton structure function.
        The curves are two different structure function parameterizations,
        the error bands reflect the uncertainties  
        due to variation of the charm mass.}

\label{fig:f2c}
\end{figure}

The statistics of about 100 events in each channel permits 
to separate the data into a small number of bins in $x$ and $Q^2$.
From the double differential cross section 
the charm contribution $\ftc$ to the proton structure 
function is extracted, 
using the the parton model formula 
\begin{equation} 
\frac{d^2\sigma^{c\bar{c}}}{dx\, dQ^2} = \frac{2\pi\alpha}{xQ^4} 
\left ( 1+(1-y)^2\right ) \cdot \ftc (x,Q^2) \;\; .
\end{equation}             
The result is shown in Fig.~\ref{fig:f2c}. 
In the probed range, $\ftc$ amounts to around 20\% 
of the total $F_2$. 
The systematic errors include those arising from the 
charm signal extraction (background shape, branching ratio)
as well as those associated with the electron measurement 
(calibration, bin center and radiative corrections). 
For comparison, NLO QCD calculations~\cite{disnlo} using 
two different structure function parameterizations
are also shown, resulting from either a flat (CTEQ2MF~\cite{cteq}) or 
steep (GRV~\cite{grv}) gluon distribution. The error bands reflect the 
uncertainties due to variation of the charm mass 
in the calculation (between 1.3 and 1.7 GeV).

$\ftc$ is defined irrespective of the production mechanism,
but the predominant mechanism is again boson gluon fusion. 
Other contributions -- scattering off a possible intrinsic charm
content of the proton, or gluon splitting into $c\bar{c}$ pairs  --
are expected to be small, as supported by  
earlier muoproduction data~\cite{emc_openc} 
and recent $e^+e^-$ results~\cite{opalgcc}.
Therefore $\ftc$ is an almost purely gluonic observable
and can provide powerful constraints on the proton structure.

\section{Conclusion}

New results in the area of hidden and open charm are emerging 
from the experiments ZEUS and H1 at HERA. 
Elastic $J/\psi$ production exhibits a very clean experimental
signature,  
and in some models it offers a temptingly high 
sensitivity to the gluon density in the proton. 
However, the theoretical discussion on how to assess the 
model uncertainties 
has not yet settled.    
Inelastic $J/\psi$ production is found in agreement 
with next-to-leading order QCD.  
The measured elasticity distribution at HERA disfavors large 
color octet contributions.
 
Charmed mesons have been measured in photoproduction 
and deep inelastic scattering.  
The total cross section $\sigma (\gamma p\ra c\bar{c})$ 
is subject to large extrapolation uncertainties. 
H1 had a first look at 
rapidity and $p_{\perp}$ distributions, 
they are in rough agreement with NLO QCD.
First evidence for diffractive charm photoproduction has been 
presented.
In DIS, a first measurement of the charm contribution
$\ftc$ to the proton structure has been made at HERA. 

Charm production at HERA thus allows to test and extend our 
understanding of heavy quark production in terms of QCD, 
and it holds the promise of providing direct determinations
of the gluon density in the proton in the near future. 
The results presented here demonstrate that this 
is progressing well along a variety of ways.
 

\subsection*{Acknowledgements}
I like to thank my colleagues from 
the ZEUS and H1 experiment for providing the 
results for this talk. 
And it is a pleasure to express my gratitude to 
Sergio Bellettini and Mario Greco for their efforts to make such a
stimulating meeting happen, 
and to their charming team to help running it so pleasantly.  
Finally, I must thank    
Antonella Bellettini and Karen Grahn 
for their unforgettable concert.

\end{document}